\documentclass[prd,aps,showpacs,nofootinbib,eqsecnum,twocolumn,groupedaddress]{revtex4}
\usepackage{amsmath} 
\usepackage{amssymb}
\usepackage{amsfonts}
\usepackage{graphicx,bm}
\usepackage{dcolumn}
\usepackage{color,amsxtra}
\usepackage{epsf}
\usepackage{enumerate}
\usepackage{hhline}
\usepackage{array}
\usepackage{tabularx}
\usepackage{subfigure}
\usepackage{appendix}
\usepackage[colorlinks=true,
            linkcolor=blue,
          urlcolor=gray,
            citecolor=red]{hyperref}

\usepackage{subfigure}

\def\be{\begin{equation}}
\def\ee{\end{equation}}
\def\bea{\begin{eqnarray}}
\def\eea{\end{eqnarray}}

\def\nn{\nonumber \\}

\begin{document}

\title{Constraining $f(T, \mathcal{T})$ gravity models using type Ia supernovae}

\author{Diego S\'aez-G\'omez${}^{1}$, C. Sofia Carvalho${}^{2,4}$, Francisco S.~N.~Lobo${}^{1,3}$ and Ismael Tereno${}^{2,3}$ } 
\affiliation{$^1$Institute of Astrophysics and Space Sciences, University of Lisbon,
\\
Edif\'icio C8, Campo Grande, 1749--016 Lisbon, Portugal}
\affiliation{$^2$Institute of Astrophysics and Space Sciences, University of Lisbon,
Tapada da Ajuda, 1349--018 Lisbon, Portugal}
\affiliation{$^3$Department of Physics, Faculty of Sciences, University of Lisbon,
\\
Edif\'icio C8, Campo Grande, 1749--016 Lisbon, Portugal}
\affiliation{$^4$Research Center for Astronomy and Applied Mathematics, 
Academy of Athens, Soranou Efessiou 4, 11--527, Athens, Greece}



\date{\today}

\begin{abstract}
We present an analysis of an $f(T, \mathcal{T})$ extension of the Teleparallel Equivalent of General Relativity, where $T$ denotes the torsion and $\mathcal{T}$ the trace of the energy-momentum tensor. This extension includes non--minimal couplings between torsion and matter. In particular, we construct two specific models that recover the usual continuity equation, namely, $f(T, \mathcal{T})=T+g(\mathcal{T})$ and $f(T, \mathcal{T})=T\times g(\mathcal{T})$. We then constrain the parameters of each model by fitting the predicted distance modulus to that measured from type Ia supernovae, and find that both models can reproduce the late--time cosmic acceleration. We also observe that one of the models satisfies well the observational constraints and yields a goodness--of--fit similar to the $\Lambda$CDM model, thus demonstrating that $f(T,\mathcal{T})$ gravity theory encompasses viable models that can be an alternative to $\Lambda$CDM. 
\end{abstract}

\maketitle

\section{Introduction}

Over the last decade, a plethora of dark energy models has been proposed in 
the attempt to understand the origin and nature of dark energy. Some theoretical proposals include scalar fields, vector fields and deviations from homogeneity \cite{DE-models}. 
A great interest has also been drawn to a theoretical framework that regards dark energy as a shortcoming of General Relativity (GR), which is the current gravity theory. In this framework, the usual way of exploring alternatives to GR has been the inclusion of extra terms in the Hilbert--Einstein action which, while preserving the 
GR predictions at local scales, introduce corrections at cosmological scales, leading 
to a late--time accelerated expansion of the Universe (for reviews on modified gravity theories, see \cite{Nojiri:2010wj}). In this framework, $f(R)$ gravity theories have been widely studied, since they appear as a natural extension of GR where the gravitational action is a more general function of the Ricci scalar $R$ that includes the Hilbert--Einstein action as a special case. From a mathematical point of view, by choosing a particular $f(R)$ action, any solution of the Friedamnn--Lema\^itre--Robertson--Walker (FLRW) equations can be generated that reproduces the desirable cosmological evolution. Most of these solutions lack viability, since they are unable to recover the GR predictions or pass the local gravity tests. Some viable $f(R)$ models of modified gravity theory were proposed that satisfy the requirements from the tests. However, they also introduce undesirable effects such as cosmological singularities. 

In addition to $f(R)$ gravity theories, other modifications of GR have been extensively explored. It is interesting to note that Einstein also introduced an alternative description of spacetime equivalent to GR, namely the Teleparallel Equivalent of General Relativity (TEGR) \cite{Unzicker:2005in} (we also refer the reader to \cite{TEGR,Teleparallelism,pereira1}). The theory is based on the assumption of the Weitzenb\"{o}ck connection instead of the Levi--Civita connection of GR, which leads to a vanishing curvature but a non--vanishing torsion $T$. This theory lies on the inertial effects that torsion produces on bodies moving within a gravitational field. In spite of the conceptual difference with respect to GR, TEGR is an equivalent description at the level of the equations. Hence, following the same reasoning, as in generalizing the Hilbert-Einstein action to a function $f(R)$,
a natural extension of TEGR consists in generalizing the TEGR action to a function $f(T)$ \cite{Ferraro:2006jd,Linder:2010py}. Moreover, since curvature and torsion differ by a total derivative, the $f(T)$ extension is not equivalent to the $f(R)$ extension. This feature guarantees that the field equations in $f(T)$ theories are second--order, which represent an advantage over the $f(R)$ theories.
As it has been extensively shown in the recent literature, by appropriately choosing the $f(T)$ action, the late--time accelerated cosmic expansion can be reproduced \cite{moddarkenergy}. In fact, $f(T)$ gravity theories 
have been explored in a wide variety of aspects, 
namely, alternative cosmological models \cite{cosmomodels}, observational constraints \cite{observational}, thermodynamics \cite{thermodynamics}, static solutions \cite{staticsols,good-and-bad,SSS_references}, extensions of other modified GR theories \cite{generalizations},
violations of the local Lorentz symmetry \cite{Li:2010cg}, existence of extra modes in gravitational waves \cite{Bamba:2013ooa}. Moreover, the viability of $f(T)$ gravity at solar system scales has been also analyzed, particularly by considering deviations from the linear action of TEGR. Such tests reveals stronger constraints on the extra terms in the action than the ones provided at cosmological scales, but keeping the aforementioned extensions of TEGR allowed by local gravity tests  \cite{Iorio:2012cm}. In addition, the junctions conditions for $f(T)$ gravity turn out more restrictive than in the Teleparallel case, but the theory still can contain the usual results of TEGR, as the Oppenheimer-Snyder collapse scenario (see \cite{delaCruz-Dombriz:2014zaa}). 

Inspired by modified gravity theory with curvature--matter couplings \cite{Bertolami:2007gv}, an extension of $f(T)$ theories that includes a non--minimal torsion--matter coupling in the action was recently explored \cite{Harko:2014sja}. The resulting theory differs both from $f(T)$ theories and from GR with a non--minimal curvature--matter coupling, providing a unified description of the expansion history. 
One can extend the models explored in \cite{Bertolami:2007gv} by considering a generalized coupling between the geometric sector and the non--geometric sector in the action, such as in $f(R,\mathcal{L}_m)$ theories,  where $\mathcal{L}_m$ is the matter Lagrangian \cite{Harko:2010mv}. Models with specific couplings between the Ricci scalar and the trace of the energy--momentum tensor $\mathcal{T}$ have also been extensively explored, such as in $f(R,\mathcal{T})$ theories \cite{Harko:2011kv}, where the dependence on $\mathcal{T}$ may be induced by exotic imperfect fluids or quantum effects (conformal anomaly), or where extra coupling terms of the form $R_{\mu\nu}T^{\mu\nu}$ can be considered \cite{Haghani:2013oma}.
However, in general, these theories imply the non--conservation of the energy--momentum tensor. In \cite{Alvarenga:2013syu}, using the standard energy--momentum conservation equation, it was shown that the matter density perturbations are different from those in the $\Lambda$CDM model, which severely constrains the viability of the models used.

Based on $f(R,\mathcal{T})$ theories, an extension of $f(T)$ allowing for a general coupling of the torsion scalar $T$ to the trace of the matter energy-momentum tensor $\mathcal{T}$ was presented in \cite{Harko:2014aja}. The resulting $f(T,\mathcal{T})$ theory is a new modification of gravity theory, since it is different from all the existing torsion or curvature based constructions. It also yields an interesting cosmological phenomenology as it encompasses a unified description of the expansion history with an initial inflationary phase, a subsequent non--accelerated, matter--dominated expansion and a final
late--time accelerating phase \cite{Momeni:2014jja}. A detailed study of the scalar
perturbations at the linear level revealed that the resulting $f(T,\mathcal{T})$ cosmology
can be free of ghosts and instabilities for a wide class of ansatzes and model parameters.

In this manuscript, we are interested in considering an $f(T,\mathcal{T})$ theory while simultaneously imposing 
the standard energy--momentum conservation equation. This provides a theoretical prior on the specific forms of the Lagrangian, analogous to that in \cite{Alvarenga:2013syu}. 
We use type Ia supernova data to further constrain the parameters of each model.

This manuscript is organized in the following manner. In Section \ref{sec2} we introduce $f(T,\mathcal{T})$ theories by presenting the general formalism. In Section \ref{sect_fTT}, we obtain the corresponding FLRW equations and construct models that satisfy the standard energy--momentum conservation equation. In Section \ref{sect_sneia}, we estimate the parameters of each model from the Union 2 type Ia supernova data compilation and compare the goodness--of--fit with the $\Lambda$CDM model. In Section \ref{sect_results} we analyse the results and in Section \ref{sect_conclusions} we present the conclusions.

\section{$f(T,\mathcal{T})$ gravity theory} \label{sec2}

We start by reviewing TEGR and its extensions. Given a manifold, the tangent space at each point $x^{\mu}$ can be described by an orthonormal basis of tetrads $e_{a}(x^{\mu})$ such that
\be
{\rm d}x^{\mu}=e_{a}^{\;\;\mu}\sigma^{a}\; , \qquad \sigma^{a}=e^{a}_{\;\;\mu}{\rm d}x^{\mu}.
\label{1.1}
\ee
Thus the line element can be written as
\begin{eqnarray}
{\rm d}s^{2} &=&g_{\mu\nu}{\rm d}x^{\mu}{\rm d}x^{\nu}=\eta_{ab}\sigma^{a}\sigma^{b}\label{1},
\label{1.1a}
\end{eqnarray} 
where 
$\eta_{ab}=\text{diag}(+1,-1,-1,-1),$ 
$e_{a}^{\;\;\mu}e^{a}_{\;\;\nu}=\delta^{\mu}_{\nu}$ and 
$e_{a}^{\;\;\mu}e^{b}_{\;\;\mu}=\delta^{b}_{a}$.  
In TEGR, instead of the Levi--Civita connection, the Weitzenb\"{o}ck connection is assumed
\begin{eqnarray}
\stackrel{\bullet}{\Gamma^{\alpha}}_{\mu\nu}=e_{a}^{\;\;\alpha}\partial_{\nu}e^{a}_{\;\;\mu}=-e^{a}_{\;\;\mu}\partial_{\nu}e_{a}^{\;\;\alpha}\label{co},
\label{WC}
\end{eqnarray}
which parallel transports the tetrads. (This is the absolute parallelism condition, where the theory takes its name from.) The Weitzenb\"{o}ck connection is not symmetric under the interchange of the lower indices, which results in a non--vanishing torsion tensor
\begin{eqnarray}
T^{\alpha}_{\;\;\mu\nu}&=&\stackrel{\bullet}{\Gamma^{\alpha}}_{\nu\mu}-\stackrel{\bullet}{\Gamma^{\alpha}}_{\mu\nu}=e_{a}^{\;\;\alpha}\left(\partial_{\mu} e^{a}_{\;\;\nu}-\partial_{\nu} e^{a}_{\;\;\mu}\right),
\label{tor}
\end{eqnarray}
which generates the gravitational field. 
Note that the Weitzenb\"{o}ck connection leads to a vanishing curvature $R^{\mu}_{\;\;\lambda\nu\rho}(\stackrel{\bullet}{\Gamma})=0$.

The difference between the Weitzenb\"{o}ck connection and the Levi--Civita connection defines the contorsion tensor
\be
K^{\alpha}_{\;\; \mu\nu}= \stackrel{\bullet}{\Gamma^{\alpha}}_{\mu\nu}-\Gamma^{\alpha}_{\mu\nu}=\frac{1}{2}\left(T_{\mu\;\;\ \nu}^{\;\; \alpha}+T_{\nu\;\;\ \mu}^{\;\; \alpha}-T_{\;\; \mu \nu}^{\alpha}\right).
\label{contor2}
\ee
The torsion scalar is defined as
\begin{eqnarray}
T&=&T^{\alpha}{}_{\mu\nu}S^{\;\;\mu\nu}_{\alpha}
      \nonumber  \\
&=&\frac{1}{4}T^{\lambda}{}_{\mu\nu}T_{\lambda}{}^{\mu\nu}+\frac{1}{2}T^{\lambda}{}_{\mu\nu}T^{\nu\mu}{}_{\lambda}-T^{\rho}{}_{\mu\rho}T^{\nu\mu}{}_{\nu},
\label{scalar-torsion}
\end{eqnarray}
where
\begin{eqnarray}
S_{\alpha}^{\;\;\mu\nu}&=&\frac{1}{2}\left( K_{\;\;\;\;\alpha}^{\mu\nu}+\delta^{\mu}_{\alpha}T^{\beta\nu}_{\;\;\;\;\beta}-\delta^{\nu}_{\alpha}T^{\beta\mu}_{\;\;\;\;\beta}\right)\label{s}.
\end{eqnarray}

The gravitational action 
for TEGR is constructed from the torsion scalar in Eq.~(\ref{scalar-torsion}), 
\begin{eqnarray}
\label{action}
S_G=\frac{1}{2\kappa^2}\int {\rm d}^4x~ e~T,
\label{TeleAction}
\end{eqnarray}
where $e=\text{det}(e^{a}{}_{\mu})=\sqrt{-g}$. The above action is equivalent to the GR action, 
which can be shown by writing the Riemann tensor as a function of the contorsion tensor in Eq.~(\ref{contor2})
\bea
&&R^{\lambda}_{\;\;\mu\rho\nu} \left(\Gamma\right)=\cr
&=&K^{\lambda}_{\;\;\mu\rho;\nu}
-K^{\lambda}_{\;\;\mu\nu;\rho}
+K^{\lambda}_{\;\;\sigma\nu}K^{\sigma}_{\;\;\mu\rho}
-K^{\lambda}_{\;\;\sigma\rho}K^{\sigma}_{\;\;\mu\nu},\qquad
\label{Riemann}
\eea
which, after contractions, yields the Ricci scalar
\be
R\,=\,-T-2D_{\mu}T^{\nu\mu}_{\;\;\;\;\;\nu}.
\label{RS}
\ee
Here $D_{\mu}$ refers to the covariant derivative defined in terms of the Levi--Civita connection. The second term on the right--hand side of Eq.~(\ref{RS}) is a total derivative that can be integrated out from the gravitational action in Eq.~(\ref{TeleAction}), thus recovering the Hilbert--Einstein action. However, teleparallel--extended $f(T)$ theories are not equivalent to GR--extended $f(R)$ theories since the total derivative in Eq.~(\ref{RS}) cannot be integrated out.

In this manuscript we study a class of theories such that the trace of the energy--momentum tensor is included in the action \cite{Harko:2014aja}
\begin{eqnarray}
S=S_G+S_m=\int {\rm d}^4x~e\left[f(T,\mathcal{T})-2\kappa^2\mathcal{L}_m\right],
\label{ftaction}
\end{eqnarray}
and the matter Lagrangian $\mathcal{L}_m$ is assumed to depend only on the 
tetrads and not on their derivatives. Here $\mathcal{T}=\mathcal{T}^{\mu}_{\;\;\mu}$ is the trace of the energy-momentum tensor, which in turn is defined by
\be
\mathcal{T}^{\mu\nu}=\frac{2}{\sqrt{-g}}\frac{\partial S_m}{\partial g_{\mu\nu}}\ .
\label{EMtensor}
\ee

Varying the action in Eq.~(\ref{ftaction}) with respect to the tetrads $e^{a}_{\;\;\mu},$ we obtain the following field equations
\begin{widetext}
\begin{eqnarray}
S^{\;\;\nu\rho}_{\mu}\left(f_{TT}\partial_{\rho}T+f_{T\mathcal{T}}\partial_{\rho}\mathcal{T}\right)+\left[e^{-1}e^{i}_{\mu}\partial_{\rho}\left(ee^{\;\;\alpha}_{i}S^{\;\;\nu\rho}_{\alpha}\right)+T^{\alpha}_{\;\;\lambda\mu}S^{\;\;\nu\lambda}_{\alpha}\right]f_{T}
+\frac{1}{4}\delta^{\nu}_{\mu}f+ \frac{1}{2} f_{\mathcal{T}}\left(\mathcal{T}^{\nu}_{\mu}+p\, \delta^{\nu}_{\mu}\right)=\frac{\kappa^2}{2}\mathcal{T}^{\nu}_{\mu}\label{em}\,,
\label{fieldEq}
\end{eqnarray}
where $f_{\mathcal{T}}=\partial{f}/\partial{\mathcal{T}}$, $f_{T}=\partial{f}/\partial{T}$ and $f_{T\mathcal{T}}=\partial^2{f}/\partial{T} \partial{\mathcal{T}}$. Using Eqs.~(\ref{Riemann}) and (\ref{RS}), the field equations in Eq.~(\ref{fieldEq}) can be rewritten 
as follows
\be
\left(R_{\mu\nu}-\frac{1}{2}g_{\mu\nu}R\right)f_{T}+\frac{1}{2}g_{\mu\nu}\left(f-f_T T\right)+2S_{\nu\mu}^{\;\;\;\;\sigma}\left(f_{TT}\nabla_{\sigma}T+f_{T\mathcal{T}}\nabla_{\sigma}\mathcal{T}\right)=\left(\kappa^2-f_{\mathcal{T}}\right)\mathcal{T}_{\mu\nu}-f_{\mathcal{T}}g_{\mu\nu}p\ .
\label{fieldEqCov}
\ee
Setting $f(T,\mathcal{T})=T$, these field equations reduce to the Einstein field equations, as expected. Moreover, since the action (\ref{ftaction}) is covariantly constructed, the field equations are divergence--free. However, by taking the divergence of the field equations (\ref{fieldEqCov}), the usual continuity equation $\nabla_{\mu}\mathcal{T}^{\mu\nu}=0$ is not recovered, but instead there results \cite{Harko:2014aja}
\be
\nabla_{\mu}\mathcal{T}^{\mu\nu}=\frac{f_{\mathcal{T}}}{\kappa^2-f_{\mathcal{T}}}\left[\left(\mathcal{T}^{\mu\nu}+g^{\mu\nu}p\right)\nabla_{\mu}\log f_{\mathcal{T}}+g^{\mu\nu}\nabla_{\mu}\left(p+\frac{1}{2}\mathcal{T}\right)\right]\ .
\label{consEq}
\ee
\end{widetext}
This is a serious shortcoming, since even in case that the extra terms of the gravitational action were important at cosmological scales only, they would affect the evolution of the different matter species in the universe. Consequently this would affect well--tested phenomenology, such as the nucleosynthesis and the photon decoupling, among others. However, as shown in the next section, some particular $f(T,\mathcal{T})$ actions can recover a divergence--free energy--momentum tensor and simultaneously lead to modifications of the FLRW equations that can reproduce the late--time cosmic acceleration with an appropriate choice of the free parameters.

\section{FLRW cosmologies in $f(T,\mathcal{T})$ gravity theory} \label{sect_fTT}

We proceed to analyse the cosmological evolution in a $f(T,\mathcal{T})$ gravity theory by assuming a spatially flat FLRW metric
\be
ds^2={\rm d}t^2-a(t)^2 \sum_{i=1}^{3}{\rm d}x_i^2\,,
\label{FLRWmetric}
\ee
where $a(t)$ is the scale factor. We assume that the matter content is described by a perfect fluid, with energy--momentum tensor $\mathcal{T}^{\mu}{}_{\nu}=\text{diag}\left(\rho_m, -p_m, -p_m, -p_m\right)$ and corresponding trace $\mathcal{T}=\rho_m-3p_m.$
Thus, the modified Friedmann equations are given by
\bea
\frac{1}{2}f-Tf_T&=&\kappa^2\rho_m-(\rho_m+p_m)f_{\mathcal{T}}\ ,  \label{FLRWeqa}
     \\
-2\dot{H}f_T-4T\dot{H}f_{TT}&=&\left(\kappa^2-f_{\mathcal{T}}\right)\left(\rho_m+p_m\right)  
   \nn
&&+2H\left(\dot{\rho}_m-3\dot{p}_m\right)f_{T\mathcal{T}}\ ,
\label{FLRWeqb}
\eea
and the divergence of the energy-momentum tensor (\ref{consEq}) leads to
\bea
&&\dot{\rho}_m+3H(\rho_m+p_m)=\cr
&=&\frac{1}{\kappa^2-f_{\mathcal{T}}}\left[\left(\rho_m+p_m\right)\partial_{t}f_{\mathcal{T}}
+f_{\mathcal{T}}\left(\dot{p}_m+\frac{1}{2}\dot{\mathcal{T}}\right)\right].\qquad
\label{consEqFLRW}
\eea
As pointed out above, the right--hand side of Eq.~(\ref{consEqFLRW}) does not in general vanish and consequently the continuity equation is not in general satisfied. 
However, by imposing the right--hand side of Eq.~(\ref{consEqFLRW}) to be zero, the usual continuity equation is recovered. 
Hence we study extended teleparallel gravity theory for two classes of actions such that the Lagrangian satisfies the condition for a divergence--free energy--momentum tensor. 

Assuming an equation of state $w_m=p_m/\rho_m$ and imposing the usual continuity equation in Eq.~(\ref{consEqFLRW})
\be
\dot{\rho}_m+3H(1+w_m)\rho_m=0\ ,
\label{ContEq}
\ee
we obtain the following condition
\be
\left(\rho_m+p_m\right)\partial_{t}f_{\mathcal{T}}+\left(\dot{p}_m+\frac{1}{2}\dot{\mathcal{T}}\right)f_{\mathcal{T}}=0\ .
\label{EqCondG}
\ee
Then, by solving Eq.~(\ref{EqCondG}) for $f(T,\mathcal{T}),$ we can obtain the corresponding gravitational action. Note that Eq. (\ref{EqCondG}) depends on the torsion scalar and its derivatives, and consequently on the cosmological background, so that in general this differential equation can not be solved exactly. However, by considering some particular forms of $f(T,\mathcal{T})$, the cosmological background, expressed through the Hubble parameter $H(t)$ 
drops out of the equation and the gravitational action can be reconstructed. 

\subsection{$f(T,\mathcal{T})=F(T)+g(\mathcal{T})$}

We first consider an action of the form $f(T,\mathcal{T})=F(T)+g(\mathcal{T})$. Then, Eq. (\ref{EqCondG}) becomes
\be
2(1+w_m)\mathcal{T}g_{\mathcal{T}\mathcal{T}}+(1-w_m)g_\mathcal{T}=0\ ,
\label{condition1}
\ee
whose general solution is given by
\be
g(\mathcal{T})=\alpha\mathcal{T}^{\frac{1+3w_m}{2(1+w_m)}}+\beta\ ,
\label{gT1}
\ee
where $\alpha$ and $\beta$ are integration constants. 
The modified Friedmann equations in Eqs.~(\ref{FLRWeqa}) and (\ref{FLRWeqb}) take the following form 
\bea
\frac{1}{2}F(T)+\frac{\alpha}{1-3w_m}\mathcal{T}^{\frac{1+3w_m}{2(1+w_m)}}+\frac{1}{2}\left(\beta-2TF_T\right) 
      \nn
-\frac{\kappa^2}{1-w_m}\mathcal{T}=0\,, \nn
\frac{1+w_m}{2(1-3w_m)}\alpha\mathcal{T}^{\frac{1+3w_m}{2(1+w_m)}}+2F_T\dot{H}+4F_{TT}\mathcal{T}\dot{H}
    \nn
+\kappa^2\frac{1+w_m}{1-3w_m}\mathcal{T}=0\,,  
\label{FLRWeq1b}
\eea
where we used $\mathcal{T}=\rho_m(1-3w)$. 
Then, by assuming a particular form of $F(T)$, we can obtain the corresponding Hubble parameter. 

For simplicity, we set $F(T)=T$ so that 
\be
f_1(T,\mathcal{T})=T+g(\mathcal{T})\ .
\label{Model1a}
\ee
This amounts to TEGR with an additive correction. We redefine the constants $\alpha$ and $\beta$ so that they become dimensionless 
\bea
\alpha&\rightarrow&\alpha\times H_0^2\left[\frac{3(1-3w_m)\Omega_mH_0^2}{\kappa^2}\right]^{-\frac{1+3w_m}{2(1+w_m)}}\,,\nn
\beta&\rightarrow& \beta\times H_0^2\,, \nonumber
\label{Variables}
\eea
where $H_0$ is the Hubble parameter today and $\Omega_m=\kappa^2\rho_{m0}/(3H_0^2)$ is the relative matter density. From here onwards, we assume that the matter content is described by dust, hence $w_m=0,$ since we are interested in late times when radiation is negligible.  
From the Friedmann equation in Eq.~(\ref{FLRWeq1b}) we derive an expression for the Hubble parameter, which we can write as a function of the redshift, $1+z=1/a,$ as follows
\be
\frac{H^2(z)}{H_0^2}=\Omega_m(1+z)^3-{\alpha\over 3} (1+z)^{3/2}-{\beta\over 6}\ .
\label{Hubble1a}
\ee
At $z=0,$ Eq.~(\ref{Hubble1a}) yields $H(0)/H_0=1,$ which entails a constraint condition for the free parameters of the model
\be
\beta=-2(3+\alpha-3\Omega_m)\ .
\label{beta1a}
\ee
Using Eq.~(\ref{beta1a}), Eq.~(\ref{Hubble1a}) becomes
\be
\frac{H^2(z)}{H_0^2}=\Omega_m(1+z)^3-{\alpha\over 3} (1+z)^{3/2}+{1\over 3}(3+\alpha-3\Omega_m)\ .
\label{Hubble1abis}
\ee
Hence, the model in Eq.~(\ref{Model1a}) contains only two free parameters, namely $\{\alpha, \Omega_m\}.$ 

\subsection{$f(T,\mathcal{T})=F(T)\times g(\mathcal{T})$}

We now consider another class of action of the form $f(T,\mathcal{T})=F(T)\times g(\mathcal{T}).$
For simplicity, we set $f(T)=T$ so that
\be
f(T,\mathcal{T})=T\times g(\mathcal{T})\ .
\label{Model2}
\ee
From Eq.~(\ref{EqCondG}) we obtain the following equation for $g(\mathcal{T})$
\be
(1+w_m)\mathcal{T}\;\partial_{t}\log\left(Tg_{\mathcal{T}}\right)+\frac{(1-w_m)}{2}\dot{\mathcal{T}}=0 \,,
\label{EqModel2}
\ee
whereas from Eq.~(\ref{FLRWeqa}) we obtain the modified Friedmann equation 
\be
\frac{1}{2}(1-3w_m)\;T\, g(\mathcal{T})+\left[\kappa^2-(1+w_m)Tg_{\mathcal{T}}(\mathcal{T})\right]\mathcal{T}=0 \,.
\label{FLRWModel2}
\ee
Note that equations (\ref{EqModel2}) and (\ref{FLRWModel2}) can be seen as a system of non-linear equations on $\mathcal{T}$ and $T$. One particular solution is given by $g(\mathcal{T})=1$, which would reduce the model to the usual TEGR gravity. By defining the variable $x=Tg_{\mathcal{T}}$, equation (\ref{EqModel2}) becomes
\be
\frac{\dot{x}}{x}=-\frac{1-w_m}{2(1+w_m)}\frac{\dot{\mathcal{T}}}{\mathcal{T}}\ ,
\label{eqNew1}
\ee
which can be integrated on both sides, leading to
\be
x(\mathcal{T})=Tg_{\mathcal{T}}=x_0\mathcal{T}^{-n}\ ,
\label{eqNew2}
\ee
where $n=(1-w_m)/[2(1+w_m)]$. Hence, by combining Eq.~(\ref{eqNew2}) with Eq.~(\ref{FLRWModel2}), the torsion scalar is dropped out and the following equation for $g(\mathcal{T})$ is obtained 
\be
\left(\frac{1}{\alpha}+\frac{1+w_m}{1-3w_m}\mathcal{T}^{-n}\right)g_{\mathcal{T}}-\frac{1}{2}\mathcal{T}^{-n-1}g=0\ .
\label{eqgT2}
\ee
The general solution to Eq.~(\ref{eqgT2}) is 
\bea
g(\mathcal{T})&=&g_0\mathcal{T}^{\frac{1-3w_m}{2(1+w_m)}}\times \cr
&\times&[(-1+3w_m)\mathcal{T}^n
-(1+w_m)\alpha]^{-\frac{1-3w_m}{2n(1+w_m)}},\qquad\quad
\label{SoleqT2}
\eea
where $g_0$ is an integration constant. By assuming $w_m=0,$ Eq.~(\ref{SoleqT2}) reduces to
\be
g(\mathcal{T})=\frac{1}{\frac{g_0}{\sqrt{\mathcal{T}}}+\frac{g_0}{\alpha}}\,,
\label{Model2G}
\ee
where we have redefined the integration constants $g_0\rightarrow \alpha/g_0$ and $\alpha\rightarrow -\alpha$ in order to make the expression for $g(\mathcal{T})$ simpler.
Then, using the Friedmann equation in Eq.~(\ref{FLRWeqa}) and redefining the free parameters in order to keep them dimensionless, i.e.
\be
g_0\rightarrow\frac{3H_0}{\kappa}g_0\,, \qquad \alpha\rightarrow \frac{H_0}{\kappa}\alpha\, ,
\label{Param3}
\ee
we find the expression for the Hubble parameter
\be
\frac{H^2(z)}{H_0^2}=g_0^2\frac{\left(\alpha+\sqrt{3\Omega_m(1+z)^3}\right)^2}{\sqrt{3}\alpha}\,.
\label{HubbleMod2a}
\ee
Note that in this case $\Lambda$CDM model is not recovered by any choice of the free parameters, due to the non-linearity of the constraint equation (\ref{EqModel2}). As with the previous model, imposing $H(0)/H_0=1$ we obtain a constraint condition for the free parameters
\be
g_0=\frac{\sqrt{3}\alpha}{(\alpha+\sqrt{3\Omega_m})^2}\ .
\label{g02}
\ee
Finally, the expression for the Hubble parameter yields
\be
\frac{H(z)}{H_0}=\frac{\alpha+\sqrt{3\Omega_m(1+z)^3}}{\alpha+\sqrt{3\Omega_m}}=\frac{1+\sqrt{3\Omega_m(1+z)^3/\alpha^2}}{1+\sqrt{3\Omega_m/\alpha^2}}\ .
\label{HubbleMod2}
\ee
Hence, this model contains only one free parameter, namely $3\Omega_m/\alpha^2.$ 

\section{Fitting $f(T,\mathcal{T})$ gravity theories to type Ia supernova data} \label{sect_sneia}

We now proceed to constrain the previous models with data. In particular, we use the Union 2 SN catalogue from \cite{Suzuki:2011hu}, which contains $N_{\text{SN}}=557$ type Ia supernovas with redshift $0.015\le z\le 1.4.$ The catalogue includes the value of the distance modulus for each SN as a function of redshift $\mu_{\text{obs}}(z)$ and the corresponding error $\sigma_{\mu_{\text{obs}}(z)}.$ 

We use the distance modulus as the observable to be fitted to the data, which is defined as
\begin{equation}
\mu_{\text{theo}}(z;\Omega_m,\alpha)=\bar\mu+5\log_{10} \left[D_L(z;\Omega_m,\alpha)\right]
\label{SN2} 
\end{equation}
Here 
$D_L (z;\Omega_m,\alpha)$ is the Hubble--free luminosity distance, 
which depends only on the model parameters $(\Omega_m,\alpha)$ as follows
\begin{equation}
D_L(z;\Omega_m,\alpha)= (1+z) \int_0^z {\rm d}z'\frac{H_0}{H(z';\Omega_m,\alpha)}\ ,
\label{SN1} 
\end{equation}
and $\bar \mu$ is an additive nuisance parameter, which contains the dependence on the Hubble parameter as follows
\bea
\bar \mu=-5\log_{10}\left[{H_0\over c}\right]+25,
\label{SN3} 
\eea
where $H_{0}$ is measured in units of $\text{km}~\text{s}^{-1}\text{Mpc}^{-1}$.

For each model, we want to find the values $(\Omega_m, \alpha)$ that best fit $\mu_{\text{theo}}$ to $\mu_{\text{obs}}.$
We start by assuming a Gaussian likelihood of the data given the models
\begin{equation}
L(\mu_{\text{obs}}(z)|\bar \mu, \Omega_m, \alpha)= {\cal N} {\rm e}^{- \chi^2/2}\ , 
\label{SN4} 
\end{equation} 
where  
\begin{eqnarray}
\chi^2 =\sum_{i=1}^{N_{\text{SN}}} \frac{(\mu_{\text{obs}}(z_i) - \mu_{\text{theo}}(z_i;{\bar \mu}, \Omega_m, \alpha))^2} {\sigma_{\mu_{\text{obs}}(z_i)}^2}\,
\label{SN5} 
\end{eqnarray}
and $\cal N$ is a normalization factor. 
In order to evaluate the likelihood, we use the Hubble function $H(z;\Omega_m,\alpha)$ for each model derived in the previous section. The nuisance parameter $\bar \mu$ is a constant offset that we can analytically marginalize and exclude from the Markov chain Monte Carlo (MCMC) sampling, so as not to increase the dimension of the parameter space. Alternatively, $\bar \mu$ can be fixed to a value. 
Following the reasoning in \cite{Leanizbarrutia:2014xta}, we expand $\chi^2$ in Eq.~(\ref{SN5}) as
\begin{equation}
\chi^2 (\Omega_m, \alpha)= A - 2{\bar \mu}B  + {\bar \mu}^2C\ ,
\label{SN6} 
\end{equation}
where the coefficients are given by
\begin{eqnarray}
A(\Omega_m, \alpha)&=&\sum_{i=1}^{N_{\text{SN}}} \frac{(\mu_{\text{obs}}(z_i) - \mu_{\text{theo}}(z_i ;{\bar \mu}=0, \Omega_m, \alpha))^2}{\sigma_{\mu_{\text{obs}} (z_i)}^2} 
\label{SN7.1} \nonumber \\
B(\Omega_m, \alpha)&=&\sum_{i=1}^{N_{\text{SN}}}\frac{(\mu_{\text{obs}}(z_i) - \mu_{\text{theo}}(z_i ;{\bar \mu}=0, \Omega_m, \alpha))}{\sigma_{\mu_{\text{obs}}(z_i)}^2} 
\label{SN7.2} \nonumber \\
C&=&\sum_{i=1}^{N_{\text{SN}}}\frac{1}{\sigma_{\mu_{\text{obs}}(z_i)}^2 } 
\label{SN7.3}
\end{eqnarray} 
Marginalizing over $\bar \mu$ amounts to considering the following $\bar \mu$--independent $\chi^2$
\bea
{\tilde\chi}^2(\Omega_m, \alpha)=A(\Omega_m, \alpha)- \frac{B(\Omega_m, \alpha)^2}{C}+\ln\left[{C\over {2\pi}}\right]
\label{SN9}
\eea
Alternatively, fixing $\bar \mu$ at the minimum of Equation (\ref{SN6}), ${\bar \mu}=B/C,$ amounts to considering the following $\bar \mu$--independent $\chi^2$
\begin{equation}
{\tilde\chi}^2(\Omega_m, \alpha)=A(\Omega_m, \alpha)- \frac{B(\Omega_m, \alpha)^2}{C}\,. 
\label{SN8}
\end{equation}


We implement the Metropolis--Hasting algorithm to produce Markov chains. We run several realizations of Markov chains, assess their convergence, remove the burn--in phase and thin them out. The following results are obtained from analysing the merged converged chains. (Yet another alternative would be to use the original expression for $\chi^2$ and run higher--dimensional chains, following the standard procedure that we used in \cite{Carvalho:2015}. The results from the two methods are fully consistent.)

\begin{figure} 
\includegraphics[width=\columnwidth]{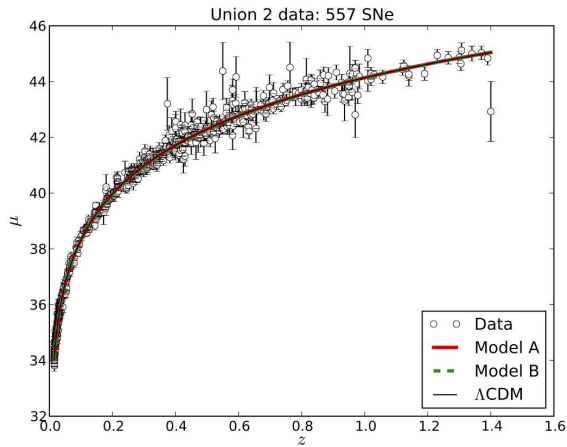}
\caption{Distance modulus as a function of redshift, including the obtained from the data points and the estimated for each model, for $H_0=70\, \text{km}~\text{s}^{-1}\text{Mpc}^{-1}$.}
\label{FigModel}
\end{figure}

For illustration in Fig.~\ref{FigModel} we plot $\mu_{\text{obs}}$ obtained from the data and the $\mu_{\text{theo}}$ computed from the values estimated for the parameters of each model from the data. We observe that the models are indistinguishable over the entire redshift range of the SN data. 

\begin{figure*} 
\begin{center}
\includegraphics[width=0.6\textwidth]{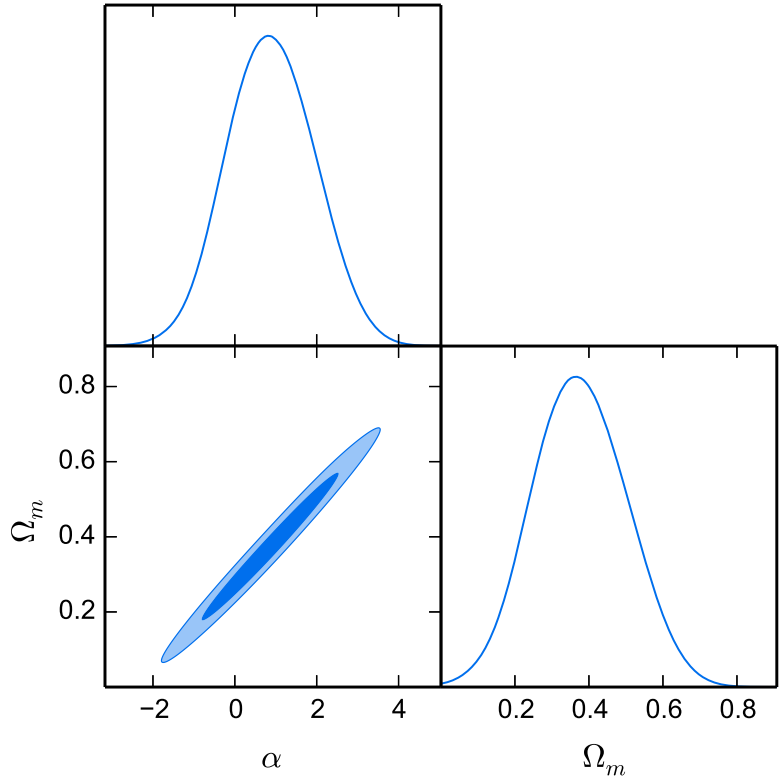}
\end{center}
\caption{Posterior distributions and contour plots for the free parameters $\{\Omega_m, \alpha\}$ of the additive model (model A).}
\label{FigModelo1a}
\end{figure*}
\begin{figure*} 
\begin{center}
\includegraphics[width=0.4\textwidth]{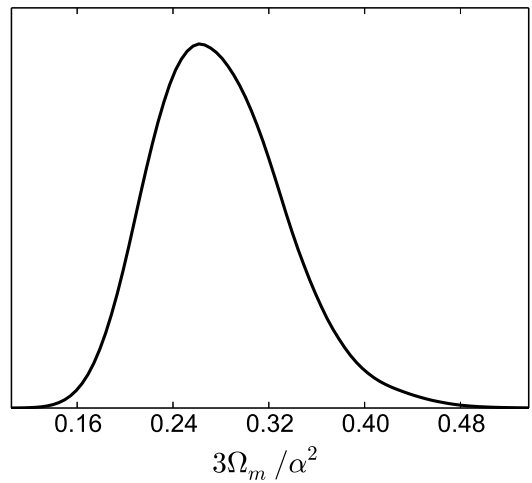}
\end{center}
\caption{Posterior probability for $3\Omega_m/\alpha^2$ of the multiplicative model (model B)}.
\label{FigModelo1b} 
\end{figure*}
 
\section{Results and Discussion} \label{sect_results}

For the additive model (model A) defined in Eq.~(\ref{Model1a}), the constraints obtained in our analysis are shown in Fig.~\ref{FigModelo1a}. 
The contour in the $(\Omega_m,\alpha)$ plane shows a very strong correlation between the two parameters, with a correlation coefficient of 0.988. Nevertheless the result is not completely degenerate and the contours do not extend indefinitely along the major axis of the ellipse. The reason for this is that, when moving along the ellipse's major axis, the points away from the central region have a negative value of $H^2$ and are thus rejected. This implies that, even though the correlation is very strong, marginalized constraints on the individual parameters are not very weak, as shown in Table~\ref{table1}, where we quote a $1\sigma$ error for $\Omega_m$ of only $41\%$ of the mean value. On the other hand, the uncertainty on $\alpha$ is quite large, allowing for both positive and negative values.

In order to try to improve these constraints we made a new MCMC analysis using Baryonic Acoustic Oscillations (BAO) data from \cite{Blake:2012pj}. We obtained weaker constraints and a contour plot with a degeneracy direction similar to that found with SN data. This did not allow us to significantly improve our estimates for model A from the combination of the two probes.

Following \cite{Tereno:2005}, we compute the principal components of our error volume from the fractional covariance matrix of the MCMC sample. The first principal component contains almost all the contribution to the error ellipse, while the second principal component defines a strong constraint along the minor axis of the ellipse, given by the equation: 
\begin{equation}
0.993\,\Omega_m-0.116\,\alpha={\rm constant}. 
\end{equation}
This is the combination of parameters that we effectively 
constrain with the data used, and that we are able to estimate with a $1\sigma$ precision of $7.5\%$ of the mean value 
\begin{equation}
\Omega_m-0.117\,\alpha=0.27 \pm 0.02\,.
\end{equation}

It is instructive to recall the Friedmann equation in Eq.~(\ref{Hubble1a}). We note that model A introduces a $\beta$ component that behaves as a cosmological constant, and an $\alpha$ component that is diluted with the scale factor as $a^{-3/2}$ as the universe expands. Assuming a constant equation--of--state for the $\alpha$ component, this would be an effective dark energy component with $w=-1/2.$ In this model, the late--time accelerated expansion would thus be driven by the behaviour derived from the $g(\cal T)$ term, added to the $F(T)$ term. We also note that the $\alpha$ component has a negative sign in Eq.~(\ref{Hubble1a}), implying that the flatness condition contributes to the same degeneracy direction as the luminosity distance observables. This indicates that, in general and contrary to what happens in $\Lambda$CDM, combining SN data with an independent probe of curvature would not break the degeneracy between the two model parameters ($\{\Omega_m,\alpha\}$ in model A and $\{\Omega_m, \Omega_\Lambda\}$ in $\Lambda$CDM). Finally, we note that model A contains $\Lambda$CDM as a particular case.

It is meaningful to address the absolute constraining power of the SN data to model A with respect to its constraining power to $\Lambda$CDM. For this we investigate the goodness--of--fit of our results for model A and of our results from an MCMC analysis of a $\Lambda$CDM model with the same number of degrees of freedom (we kept the same number of free parameters by not imposing flatness on the $\Lambda$CDM model). For this purpose, we show in Table~\ref{table1} the best--fit $\chi^2$ and reduced $\chi^2$ values, as well as the Deviance Information Criterium. The reduced $\chi^2_{\rm red}$ is defined as
\be
\chi^2_{\rm red} =\dfrac{\chi^2_{\rm min}}{N-d-1},
\ee
where $N$ is the number of experimental points used ($N_{\text{SN}}$) and $d$ is the number of parameters of the model. The Deviance Information Criterium (DIC) \cite{DIC} provides a way of penalizing a model according to the complexity of its posterior probability distribution. Here we compute it as
\be
DIC=\chi^2_{\rm min} + 2 p_D
\ee
where $p_D=\bar\chi^2-\chi^2_{\rm min}$ is called the Bayesian complexity and strongly depends on the average of the $\chi^2$ values of the sample. We find that the goodness--of--fit of model A is almost identical to that of $\Lambda$CDM, even being less penalized when taking into account the complexity of its posterior distribution.


\begin{table*}
\begin{center}
\begin{minipage}{0.88\textwidth}
\caption{$1\sigma$ SN estimates for the parameters of the spatially flat models A, B and flat $\Lambda$CDM model and also for the non--flat $\Lambda$CDM. The best--fit $\chi^2$, Bayesian complexity and DIC values are also shown for the three cases.
\label{table1}}
\end{minipage}\\
\vspace{0.2cm}
\begin{tabular}{cccccc}
\hline
\hline
\bf{Model} & MCMC parameters   & $\bf{\chi_{\rm min}^2}$ & $\bf{\chi_{\rm red}^2}$ & $p_D$ & $DIC$\\
\hline \vspace{-5pt}\\
$\Lambda$CDM  & $\Omega_m= 0.29 \pm 0.08$\ ,\ $\Omega_\Lambda= 0.76 \pm 0.12$ & 550.53 & $0.99 $ & 2.0 & 554.58\\
\\
Model A & $\Omega_m= 0.32 \pm 0.13$\ ,\  $\alpha=0.45\pm  1.13$ & $550.56$  & 0.99 & 1.8 & 554.17\\
\\
Model B & $3\Omega_m/\alpha^2=0.27\pm0.05$ & $552.57$ & 0.99 & $1.0$ & 554.57\\
\\
flat $\Lambda$CDM  & $\Omega_m= 0.27 \pm 0.02$\ ,\ $\Omega_\Lambda= 0.73 \pm 0.02$ & 550.66 & $0.99 $ & 1.0 & 552.65\\
 \hline \hline
\end{tabular}
\end{center}
\end{table*}
 

We turn now to the multiplicative model (model B) defined in Eq.~(\ref{Model2}). Contrary to the additive model, it was possible to write the Hubble function in terms of a single parameter, $3\Omega_m/\alpha^2$, which is a ratio of the two original parameters. This shows explicitly that, as before, the luminosity distance is a function of a combination of the two free parameters. However, in this case, we can produce Markov chains in terms of the single parameter from the start. 

The constraint obtained from our analysis of SN data is shown in Fig.~\ref{FigModelo1b}, where the posterior probability distribution of the ratio parameter is depicted. The $1\sigma$ constraint on the ratio parameter is around $20\%$ of the mean value, which is weaker than the constraint found for model A. In Table~\ref{table1} we show the result and the goodness--of--fit statistics.
The goodness--of--fit, as given by the best--fit, is worse than in model A, but we find a smaller penalizing factor, implying a DIC value also similar to $\Lambda$CDM.

\section{Conclusions}\label{sect_conclusions}

In this manuscript, we analysed the viability of $f(T,\mathcal{T})$ gravity theories, which are extensions of TEGR. 
These theories contain a non--minimal coupling between curvature and matter, and in general do not satisfy the usual continuity equation. 
By imposing the condition for a divergence--free energy--momentum tensor, we found two models, namely an additive and a multiplicative extension of TEGR.

We computed the cosmological evolution predicted by each model and constrained the free parameters by fitting the theoretical distance modulus to that measured from the type Ia supernovae compiled in the Union 2 catalogue. Both models contain two and one free parameter respectively. We found that both models can reproduce late--time cosmic acceleration. In both models, the data allowed us to estimate with good precision a combination of the models' parameters, namely
\be
\Omega_m-0.12\,\alpha=0.27 \pm 0.02 \,
\ee
for the additive model (\ref{Model1a})
and
\be
3\Omega_m/\alpha^2=0.27\pm0.05 \,
\ee
for the multiplicative model (\ref{Model2}). By comparing the goodness--of--fit and the DIC values, we found that model A is 
equally well constrained by SN data as $\Lambda$CDM. This model has one extra parameter and does not introduce extra complexity when compared to the non--flat $\Lambda$CDM. Indeed they both have similar $p_D$ values. We note that the additive model contains $\Lambda$CDM as a particular case, which is not the best fit although it is within the 1$\sigma$ region. Model B, with just one parameter, has similar complexity as the flat $\Lambda$CDM. However, its goodness--of--fit is worse, implying a higher DIC value.

Hence, we have showed that $f(T,\mathcal{T})$ gravity theory encompasses viable models that can be considered as an alternative to the $\Lambda$CDM model at the background level. On the other hand, previous works on the behaviour of extended theories of TEGR suggest the viability of such theories when confronting with solar system tests, where TEGR is recovered \cite{Iorio:2012cm}. In the same way, the class of $f(T, \mathcal{T})$ theories analyzed in this manuscript should also recover TEGR at local scales. While this is an essential feature of every modification of TEGR, an exhaustive analysis, similar to the ones done within the framework of other modified gravities (see for instance \cite{Capozziello:2007ms}), will require full attention of a future work.

%

\begin{acknowledgments}
This work was supported by Funda\c{c}\~ao para a Ci\^encia e a Tecnologia (FCT) through the research grant UID/FIS/04434/2013. 
CSC is funded by FCT, 
Grant No.~SFRH/BPD/65993/2009. 
FSNL acknowledges financial  support from FCT  
through the Investigador FCT Contract No.~IF/00859/2012, funded by FCT/MCTES
(Portugal).
DSG acknowledges support from FCT, 
Grant No.~SFRH/BPD/95939/2013.
IT acknowledges support from FCT 
through the Investigador FCT Contract No.~IF/01518/2014 and POPH/FSE (EC) by FEDER funding through the program Programa Operacional de Factores de Competitividade -- COMPETE.
\end{acknowledgments}



\end{document}